\documentclass[twocolumn]{aastex63}
\usepackage{multirow}
\usepackage{amsmath}
\usepackage{booktabs}

\usepackage{xcolor}

\newcommand\lsim{\mathrel{\rlap{\lower4pt\hbox{\hskip1pt$\sim$}}
\raise1pt\hbox{$<$}}}
\accepted{ApJL}
\shorttitle{IMF of DM-Free Gas Objects}
\shortauthors{Lake et al.}
\graphicspath{{./}{figures/}}

\begin{document}

\title{The Stellar Initial Mass Function of Early Dark Matter-free Gas Objects}


\correspondingauthor{William Lake}
\email{wlake@astro.ucla.edu}
\author[0000-0002-4227-7919]{William Lake}
\affil{Department of Physics and Astronomy, UCLA, Los Angeles, CA 90095, USA\\}
\affil{Mani L. Bhaumik Institute for Theoretical Physics, Department of Physics and Astronomy, UCLA, Los Angeles, CA 90095, USA\\}
\affil{Department of Physics and Astronomy, Dartmouth College, Hanover, NH 03755, USA \\}

\author[0000-0002-1655-5604]{Michael Y. Grudić}
\affiliation{Center for Computational Astrophysics, Flatiron Institute, 162 Fifth Avenue, New York, NY 10010, USA \\}

\author[0000-0002-9802-9279]{Smadar Naoz}
\affil{Department of Physics and Astronomy, UCLA, Los Angeles, CA 90095}
\affil{Mani L. Bhaumik Institute for Theoretical Physics, Department of Physics and Astronomy, UCLA, Los Angeles, CA 90095, USA\\}

\author[0000-0001-7925-238X]{Naoki Yoshida}
\affiliation{Department of Physics, The University of Tokyo, 7-3-1 Hongo, Bunkyo, Tokyo 113-0033, Japan}
\affiliation{Kavli Institute for the Physics and Mathematics of the Universe (WPI), UT Institute for Advanced Study, The University of Tokyo, Kashiwa, Chiba 277-8583, Japan}
\affiliation{Research Center for the Early Universe, School of Science, The University of Tokyo, 7-3-1 Hongo, Bunkyo, Tokyo 113-0033, Japan}

\author[0000-0003-2369-2911]{Claire E. Williams}
\affil{Department of Physics and Astronomy, UCLA, Los Angeles, CA 90095}
\affil{Mani L. Bhaumik Institute for Theoretical Physics, Department of Physics and Astronomy, UCLA, Los Angeles, CA 90095, USA\\}

\author[0000-0001-5817-5944]{Blakesley Burkhart}
\affiliation{Department of Physics and Astronomy, Rutgers, The State University of New Jersey, 136 Frelinghuysen Rd, Piscataway, NJ 08854, USA \\}
\affiliation{Center for Computational Astrophysics, Flatiron Institute, 162 Fifth Avenue, New York, NY 10010, USA \\}

\author[0000-0003-3816-7028]{Federico Marinacci}
\affiliation{Department of Physics \& Astronomy ``Augusto Righi", University of Bologna, via Gobetti 93/2, 40129 Bologna, Italy\\}
\affiliation{INAF, Astrophysics and Space Science Observatory Bologna, Via P. Gobetti 93/3, 40129 Bologna, Italy\\}

\author[0000-0001-8593-7692]{Mark Vogelsberger}
\affil{Department of Physics and Kavli Institute for Astrophysics and Space Research, Massachusetts Institute of Technology, Cambridge, MA 02139, USA\\}

\author[0000-0002-8859-7790]{Avi Chen}
\affiliation{Department of Physics and Astronomy, Rutgers, The State University of New Jersey, 136 Frelinghuysen Rd, Piscataway, NJ 08854, USA \\}



\begin{abstract}
 Among the remarkable strides made by JWST is the discovery of the earliest star clusters found to date. These have been proposed as early progenitors of globular clusters, which are known to come from the early stages of star formation in the Universe. This is an exciting development in modern astronomy, as it offers an opportunity to connect theoretical models of globular cluster formation to actual observations of these high-redshift structures. In this work, we aim to develop observational signatures of a star cluster formation route known as supersonically induced gas objects, which are dark matter-less gas clouds in the early Universe proposed as a potential origin of some globular clusters. For the first time, we follow the star formation process of these early Universe objects using high-resolution hydrodynamical simulations, including mechanical feedback. Our results suggest that the first dark matter-less star clusters are top-heavy, meaning that they have a flatter IMF slope compared to very young low-metallicity star clusters in the local Universe, and they also have extremely high stellar mass surface densities compared to their local counterparts.
\end{abstract}


\section{Introduction}\label{sec:main}


To date, JWST has detected the earliest known star clusters in our Universe \citep[e.g.,][]{Adamo+24,Messa+24,Vanzella+24, Mowla+24}.  They appear to be relatively compact \citep[$\sim$~few~pc, e.g.,][]{Adamo+24} and had only recently formed their stars. It was speculated that these clusters may be the earliest progenitors of globular clusters ever detected.
Globular clusters are a relic of the initial stages of star formation in the Universe. However, because they contain little to no dark matter \citep[e.g.,][]{Heggie+96,Bradford+11,Conroy+11,Ibata+13}, their formation mechanism poses a significant theoretical challenge. 

A recent suggestion pointed out that the relative velocity between the gas and the dark matter \citep{TH} in the early Universe could naturally form potentially star-forming regions outside of dark matter halos.
This relative velocity arose well before Recombination, when dark matter fell towards primordial inhomogeneities, allowing them to grow over time to form the seeds of the galaxies we see today. Until Recombination, the gas was tightly coupled to radiation, which suppressed the growth of gas overdensities \citep{NB}. 
Because the motion of baryons was suppressed by the radiation field, the infall of the dark matter towards overdense regions gave rise to baryon-dark matter relative motions, known as baryon-dark matter streaming or as the streaming velocity \citep{TH,tseliakhovich11}. These motions are highly supersonic ($\sim30$~km s$^{-1}$ at $z=1100$, or about $5$ times the speed of sound) and generate spatial offsets between gas clumps and the early dark matter overdensities that give rise to them, depleting the newborn clumps of dark matter \citep{naoz12,naoznarayan14}. The gas clumps, devoid of dark matter (like globular clusters, or GCs), form baryon-enriched structures with masses below $10^7$~M$_\odot$ known as supersonically induced gas objects (SIGOs), which have been shown to form star clusters through simulations and theory \citep{popa,chiou18,chiou+19,Chiou+21,Lake+21,Williams+22,Lake+22,Nakazato+22,Lake+24}. These objects form in similar abundance to low-metallicity globular clusters in the early Universe \citep{Lake+21}, and because the Milky Way may have formed in a region of the Universe with an elevated value of the streaming velocity \citep{Uysal+22}, the local Universe is expected to have formed a higher abundance of SIGOs.

\begin{figure*}[]
\centering
\includegraphics[width=0.9\textwidth]{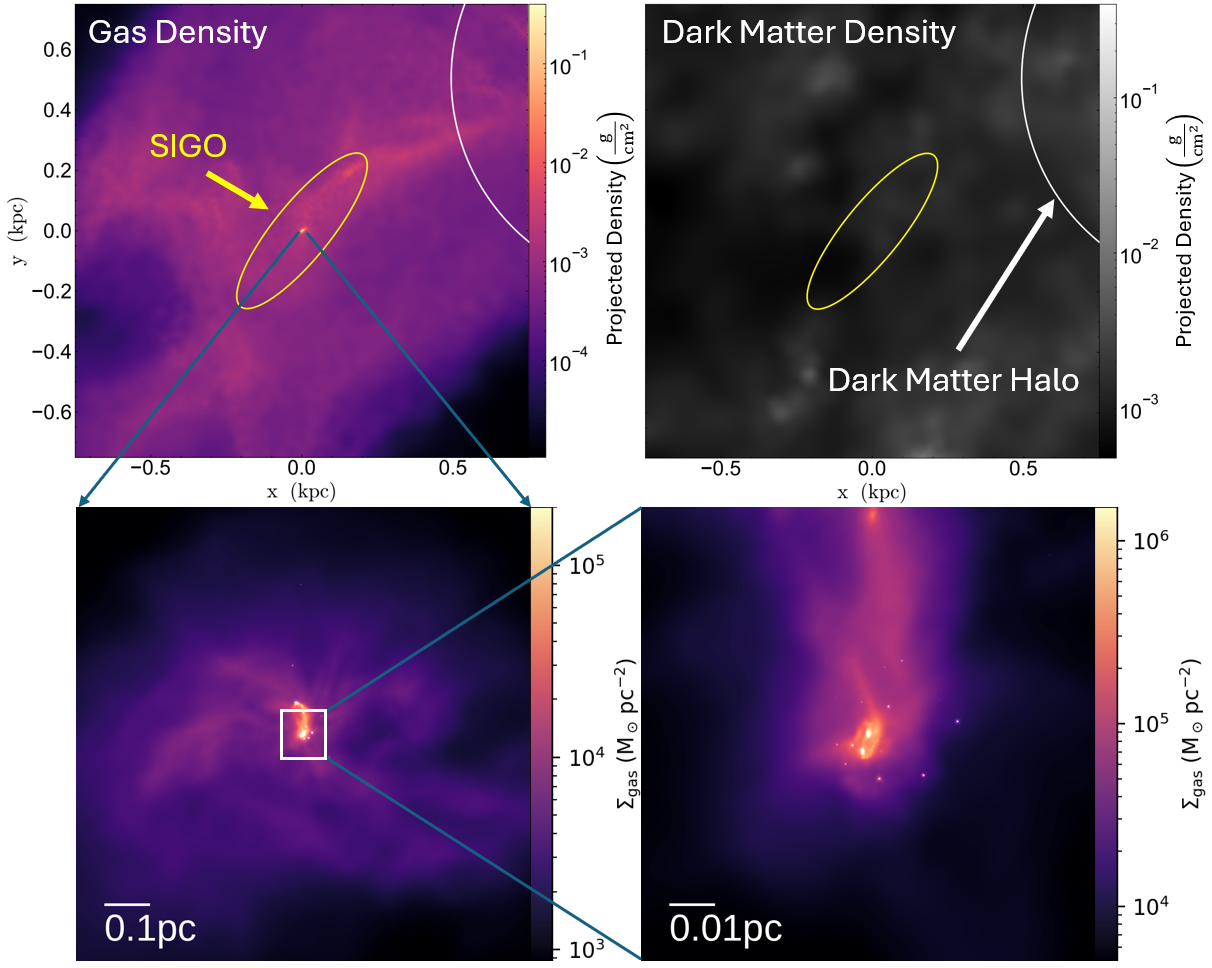}
\caption{A star-forming SIGO (yellow) in {\tt AREPO} and {\tt STARFORGE} simulations. In the top line ({\tt AREPO} simulations), it is near a DM halo (white). The left top panel shows the local gas density, while the right top panel shows the local DM density (and lack thereof, in the SIGO). The bottom panels show a similar SIGO (i.e., not a zoom simulation from the top panel) in a metal-poor {\tt STARFORGE} simulation, with much higher resolutions and on much smaller scales.}\label{fig:arepo}
\end{figure*}

A model of such an object is shown in the top panels of Figure~\ref{fig:arepo}. Specifically, we show a SIGO extracted from a cosmological simulation using {\tt AREPO} \citep{springel10} at $z=15$. The top left panel shows a self-gravitating gas-dominated structure at kpc scales, while the right shows the lack of a DM component in the same object\footnote{See Methods~\ref{ssec:Arepo} for simulation details}. This depiction of a SIGO shows it just before star formation begins at its core.

Studies of these SIGOs in Milky Way-scale cosmological simulations have determined their abundances \citep{popa,Lake+21,Lake+22}, some observational signatures such as their sizes and initial stellar mass estimates \citep{chiou+19,Lake+23,Lake+24,Williams+24}, and some population-level properties such as their mass distributions and spin distributions \citep{chiou18,Chiou+21,Lake+21,Lake+22,Lake+23,Lake+24, Williams+24}. However, all these simulations studied SIGOs at the inherently limited resolution required for Mpc-scale simulations. This restricted their ability to probe the interior of the SIGOs, as the resolution of these runs ($\sim10^2$~M$_\odot$) is not sufficient to resolve the objects' cores. This level of resolution also artificially limited the effect of molecular hydrogen chemistry in the cooling and subsequent gravitational collapse of  SIGOs, as well as under-resolving the gravity and hydrodynamics. One way to avoid these numerical issues is by conducting higher-resolution simulations based on cut-out simulation volumes \citep{Nakazato+22}. Yet, explicitly modelling star formation that follows the gravitational collapse of SIGOs to form stars remains a challenge even using the improved resolution of the aforementioned numerical methodology. Moreover, none of these studies included feedback, and none probed the stellar properties of SIGOs (i.e., the Initial Mass Function, IMF). This limited the interpretation of the results, leaving open key questions such as the star formation efficiency of SIGOs and their ultimate luminosities and observability by JWST and future instruments.

Here, guided by the properties of the star-forming SIGOs in our lower-resolution cosmological simulations \citep{Lake+23}, we investigate the star-formation process at substantially higher resolutions. We utilize small-scale, very-high-resolution ($\sim 10^{-2}$~M$_\odot$) {\tt STARFORGE} simulations \citep{Grudic+21} performed with the {\tt GIZMO} code. These simulations resolve the evolution of a single SIGO analogue through its cooling, collapse, and resolved star formation. This allows us to determine the properties of the SIGO, i.e., the first dark matter-free star cluster. Specifically, we ascertain its IMF and stellar mass, and estimate its star formation efficiency.


\begin{table*}[t]\label{Table:Runs}
\centering
\hspace*{-.125\linewidth}
\begin{tabular}{|c|c|c|c|c|c|c|c|c|}
\hline
Name & Metallicity (Z$_\odot$) & Jets? & $R_{\rm eff}$ (pc) & $\Sigma$ (M$_\odot$ pc$^{-2}$) & M (M$_\odot$)  & M$_{\rm gas, init}$ (M$_\odot$) & R$_{\rm gas, init}$ (pc) & T$_{\rm gas, init}$ (K) \\ \hline
M6NJ & $10^{-6}$               & No  & $0.025$ & $1.5\times10^5$ & $590$ & $3\times10^5$ & 130 & 225  \\ \hline
M6J  & $10^{-6}$               & Yes  & $0.19$ & $1.0\times10^3$ & $230$ & $3\times10^5$ & 130 & 225 \\ \hline
M4NJ & $10^{-4}$               & No   & $0.05$ & $3.7\times10^4$ & $580$ & $3\times10^5$ & 130 & 225 \\ \hline
M4J  & $10^{-4}$               & Yes & $0.03$  & $4.9\times10^4$ & $280$ & $3\times10^5$ & 130 & 225 \\ \hline
\end{tabular}
\caption{Description of simulation runs performed. Runs have differing metallicities and either contain or neglect jet feedback. Names for different runs are given. Effective radii are calculated as the half-light radius $100$~kyr after the formation of the first star, and stellar surface densities are calculated using the same effective radii.}
\end{table*}

\section{Methods}\label{Sec:Methods}

\subsection{AREPO}\label{ssec:Arepo}

Our high-resolution {\tt STARFORGE} simulations of SIGOs are inspired by a set of small-box ($2.5$~cMpc) {\tt AREPO} simulations \citep{springel10} performed and analyzed in \citet{Lake+23} and \citet{Lake+24}. These simulations included $768^3$ DM particles with mass M$_{\rm DM} = 1.1 \times 10^3 $ M$_\odot$, and $768^3$ Voronoi mesh cells with gas mass M$_{\rm B} = 200$ M$_\odot$, and were run from $z=200$ to $z=12$. Examples of a SIGO from this simulation are shown in the top row of Figure~\ref{fig:arepo}, showing both the gas overdensity and the lack of a corresponding dark matter overdensity, as well as showing in rough terms the limited resolution of the simulation (which is capable of resolving the overall SIGO but not the central star cluster). Initial conditions for these simulations used transfer functions from a modified version of {\tt CMBFAST} \citep{seljak96}, which incorporates first-order scale-dependent temperature fluctuations \citep{NB} and the streaming velocity \citep{TH}. This creates separate transfer functions for dark matter and baryons at $z=200$ that account for the suppression of structure from the streaming velocity \citep{popa}. The streaming velocity was implemented as a uniform velocity boost along the x-axis based on its $2\sigma$ value at $z=200$ of $11.8$~km~s$^{-1}$. The simulations also included non-equilibrium molecular hydrogen cooling based on {\tt GRACKLE} \citep{Smith+17, Chiaki+19}. This package includes non-equilibrium molecular hydrogen and deuterium chemistry. However, it does not include metal line cooling, nor does it include external radiation such as Lyman-Werner backgrounds. 

\subsection{STARFORGE}\label{ssec:Starforge}

As discussed above, the high-resolution simulations utilize the {\tt STARFORGE} star formation framework within the {\tt GIZMO} hydrodynamic simulation code \citep{Hopkins+15,Grudic+21}. These simulations include metal and molecular hydrogen cooling, molecular hydrogen nonequilibrium chemistry, and external heating from an interstellar radiation field (ISRF). The cooling and chemistry is based on FIRE-3 \citep{Hopkins+23} and takes into account recombination, thermal bremsstrahlung, metal lines, molecular lines, fine structure, and dust collisional processes. These processes differ from the {\tt AREPO} simulations above in two main ways. Firstly, the {\tt AREPO} simulations contain no metal chemistry or cooling, but they contain an explicit tracing of deuterium species abundances. Secondly, the {\tt AREPO} simulations do not contain an ISRF heating term, as they are run for much longer time periods and scales over which the ISRF varies more. In the bottom panels of Figure~\ref{fig:arepo} we illustrate the resolution achieved in this type of simulation, which is $0.01$~M$_\odot$. As discussed in Appendix~\ref{ssec:Convergence}, this resolution is well-converged with respect to the mass functions discussed in this paper. We display the star-forming region of the SIGO $13$~kyr after the formation of the first star, from our simulation with $10^{-6}$~Z$_\odot$ metallicity without jet feedback (run M6NJ, see also Table \ref{Table:Runs}), and compare it to the resolution achieved in the {\tt AREPO} simulations that these initial conditions are based on. As shown, the simulation is well-resolved on scales of hundredths of a parsec.

These {\tt STARFORGE} simulations also explicitly model feedback mechanisms through supernovae and protostellar jets. As we are only simulating these clouds for about $100$~kyr after the formation of the first star, no supernovae occur in our simulations. The existence and strength of jets in early, low-metallicity protostars is currently under debate \citep{Machida+08,Machida+13,Sharda+20,Sadanari+21,Prole+22,Sadanari+24}. Thus, for each run at different metallicities, we perform two runs, where the jets are ``on'' or ``off.'' We model the protostellar jets using the mechanism described in \citet{Grudic+21}, launching them with a velocity equal to $30\%$ that of the keplerian velocity at the protostellar radius and a mass equal to $30\%$ of the accreted mass by the protostar. Because of our extremely low metallicities, stellar winds are not significant for star formation regulation in these clouds \citep[see for a review][]{Klessen+23}, and we neglect them. Radiative feedback is likely to have an effect on long timescales in these early, low-metallicity clouds by driving gas outflows and ionization \citep[e.g.][]{Stacy+12,Guszejnov+22}. However, the exact impact on the short timescales simulated here is uncertain due to the unknown extent to which radiation can escape the high-density accretion regions surrounding Pop III stars \citep[see, e.g.][]{Jaura+22, Sharda+24}. The effect of radiative feedback on our timescales is beyond the scope of this paper and is not included.

Star formation is modeled according to the set of prescriptions implemented in {\tt STARFORGE} discussed in  \citet{Grudic+21} and reproduced here. To be considered for star formation and become a sink particle, a gas cell must first satisfy a minimum density criterion based on the Jeans scale: in our simulations with a resolution of $0.01$~M$_\odot$, this density can be expressed as $\rho_J\approx4.7\times10^{-12}\,(c_s/{\rm km~s^{-1}})^6$~g~cm$^{-3}$, where c$_s$ is the local speed of sound. Secondly, the gas cell must be the densest cell of the cells with overlapping kernel radii (for this purpose, existing sinks have infinite density). Thirdly, the gas cell's density must be increasing with time. Fourthly, the cell must be gravitationally unstable at the resolution scale, determined by a local Jeans analysis including contributions from the thermal pressure and velocity dispersion \citep{Federrath+10, Hopkins+13}. This permits star formation given the following virial condition:
\begin{equation}
\alpha_g = \frac{\frac{2\pi^2}{\Delta x^2}c_s^2+||\mathbf{\nabla \mathit{v}}||^2}{4\pi G\rho} < 2,
\end{equation}
where $\Delta x = (\Delta m/\rho)^{1/3}$ is the local cell length, and ‖ · ‖ is the Frobenius norm. Fifth, the tidal tensor at the position of the gas cell must be fully compressive. Finally, the free-fall timescale for the gas cell must be shorter than both the free-fall and orbital timescales to nearby sink particles.

Upon becoming a sink particle, the new star can accrete gas particles based on the conditions detailed in \citet{Grudic+21}. Namely, a gas cell is eligible for accretion based on the satisfaction of four criteria. First, the center of the gas cell must be within the sink radius. Second, the gas cell must be gravitationally bound to the sink. Third, the gas cell's angular momentum relative to the sink must be less than that of a circular Keplerian orbit at the distance of the gas cell from the sink. Fourth, the volume of the gas cell must be less than the spherical volume described by the sink radius, preventing unresolved gas cells from accreting onto the sink. In cases where the gas cell is eligible for accretion by multiple sinks, it is accreted by the sink with the shortest mutual dynamical time $t_{\rm dyn} = \sqrt{\frac{r_{gs}^3}{G(m_g + m_s)}}$, where $r_{gs}$ is the gas-sink separation and $m_g$ and $m_s$ are the gas cell and sink masses.

\subsection{Initial Conditions}\label{ssec:IC}

Our simulations are a proof-of-concept motivated by the properties of SIGOs in {\tt AREPO} simulations \citep{Lake+23}. The simulations begin with a uniform sphere of gas\footnote{At formation, SIGOs often do not have a significantly non-uniform density profile in simulations, so this was chosen for simplicity.} with a mass of $3\times10^5$ M$_\odot$ generated using {\tt MakeCloud} \citep{MakeCloud}, representing a SIGO with sufficient mass to form stars, but with a relatively low mass for such an object. This was chosen for computational feasibility, allowing us to explore several cloud setups with different feedback and metallicity prescriptions. The cloud has a radius $R_{\rm cloud} = 130$~pc and an initial temperature of $225$~K, consistent with molecular hydrogen cooling in pristine gas and with SIGOs in our {\tt AREPO} simulations. The mass and temperature here are chosen as the mass and mean temperature of the SIGO presented in \citet{Lake+23}, and the radius is chosen to describe a sphere of equivalent volume to the SIGO--approximately $0.009$~kpc$^3$--for simplicity (though note that SIGOs can exist as eccentric ellipsoids, the effect of which on star formation is beyond the scope of this letter). The SIGO is initialized with a spin parameter of $\lambda_S = 0.01$, which is representative of the gas in the SIGO from \citet{Lake+23} when star formation began and broadly typical of SIGOs \citep{chiou18}. Here, the spin parameter is defined as \begin{equation}
    \lambda_{\rm S} = \frac{J_g}{\sqrt{2}M_gv_{c}R_{\rm max}},
\end{equation}
where $J_g$ is the angular momentum of the gas, $M_g$ is the total gas mass in the SIGO, $v_{c}$ is the circular velocity at a distance $R_{\rm max}$ from the center of the ellipsoid, and $R_{\rm max}$ is the maximum axis length of the SIGO determined by an ellipsoid fit. The cloud is surrounded by warm, diffuse gas to ensure that the SIGO is in thermal pressure equilibrium with its surroundings. The cloud's initial turbulent energy is specified using $\alpha_{\rm turb}$ \citep{Bertoldi+92}:
\begin{equation}
\alpha_{\rm turb} = \frac{5||v_{\rm turb}||^2R_{\rm cloud}}{3{\rm GM}},
\end{equation}
where $v_{\rm turb}$ is the turbulent velocity field, which is initialized using a Gaussian random field with a $k^{-2}$ power spectrum, which is a representative power spectrum of the SIGOs in the simulations of \citet{Lake+24}. We take $\alpha_{\rm turb} = 0.5$, which is approximately its median value for star-forming SIGOs in our reference {\tt AREPO} simulations \citep[see][]{Lake+23} and also a commonly used value for other star formation models \citep{Burkhart_2021,Appel_2022}. 

The early star formation process is sensitive to the gas composition. Metals enable more efficient cooling, facilitating gravitational collapse and star formation. SIGOs do not evolve in isolation, and so we must consider whether nearby supernovae can impact their gas metallicities. More particularly, large protogalaxies can host SIGOs \citep{Lake+23}, and thus, star formation in host galaxies may contribute small amounts of metals to the SIGO. 
This may ``pollute'' the SIGO with metals, allowing it to form stars from gas that is not truly pristine \citep[see][]{Schauer+21}. Therefore, here we consider two scenarios for external metal pollution: one with low levels of metals (Z$_{\rm SIGO}=10^{-6}$~Z$_\odot$) in which metallicity effects are minimal \citep{Omukai+10}, and one with relatively high levels of metal enrichment that take the SIGO to the boundaries of Pop III to Pop II star formation (Z$_{\rm SIGO}=10^{-4}$~Z$\odot$). The primary coolant in both cases, regardless, is molecular gas, which we trace explicitly using the methods of \citet{Hopkins+23}. We instantiate the cloud with a molecular hydrogen mass fraction of $10^{-3}$ throughout, corresponding to the median value observed in star-forming SIGOs in \citet{Lake+23} simulations, and typical of early star-forming molecular clouds \citep{Klessen+23}. The ambient radiation field is chosen to mimic a $z=20$ radiation background for the mean Universe. Table~\ref{Table:Runs} reports the key properties for the full set of simulations carried out in this work.

\section{Results}\label{Sec:results}

\begin{figure*}[ht]
\centering
\includegraphics[width=0.28\textwidth]{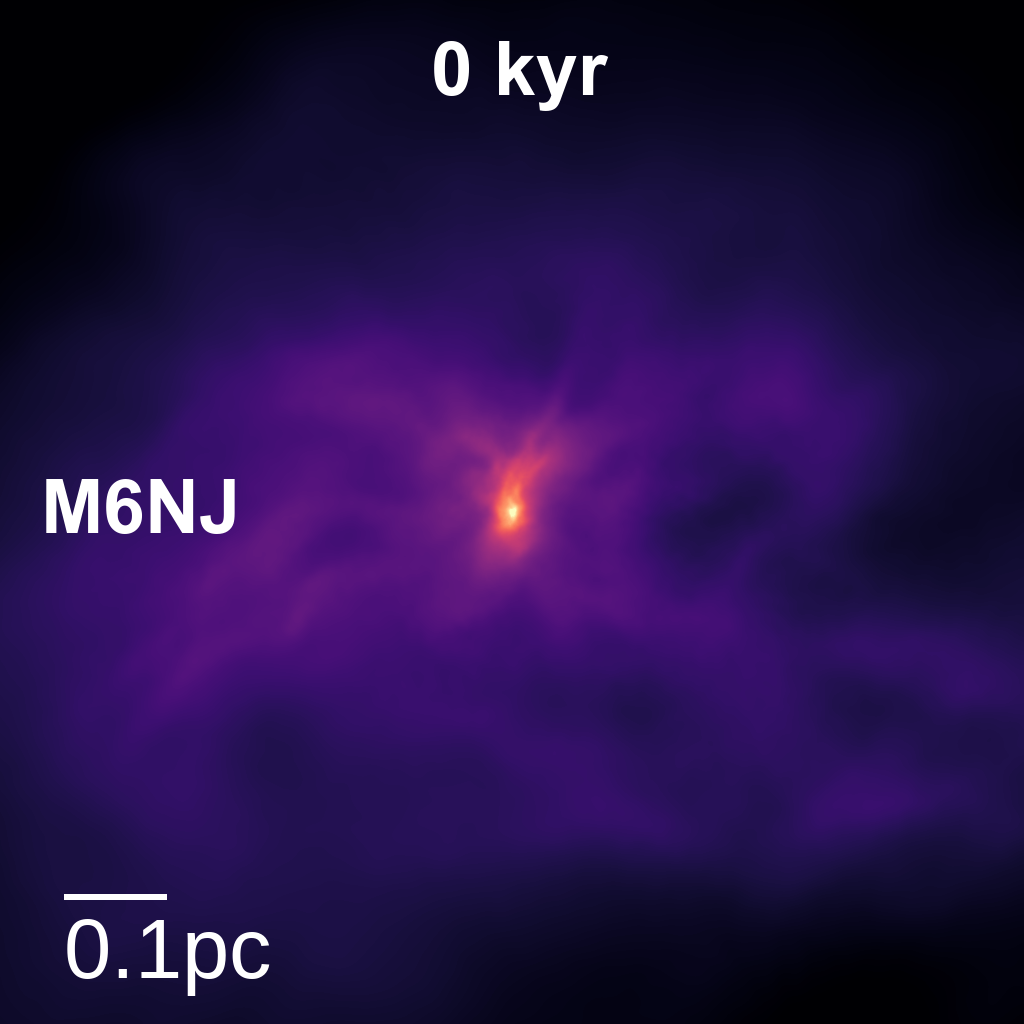}
\includegraphics[width=0.28\textwidth]{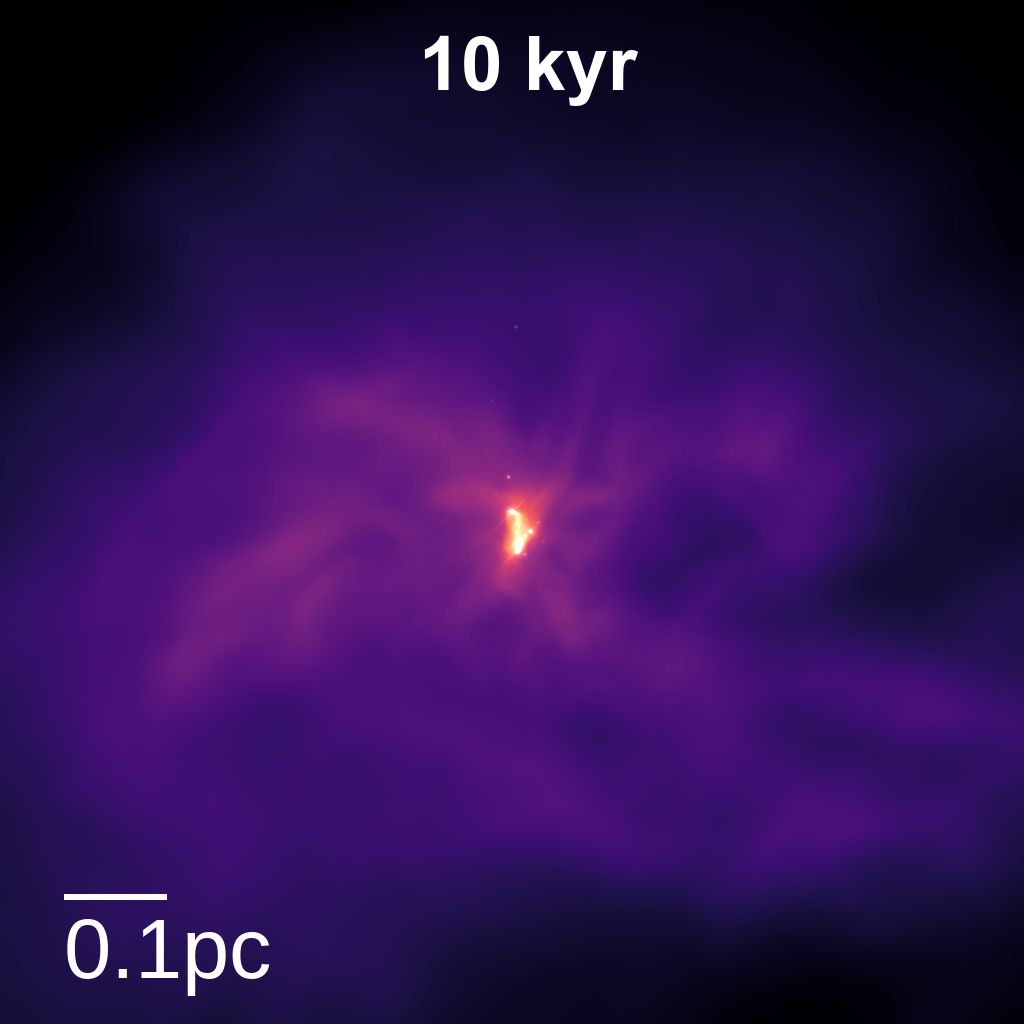}
\includegraphics[width=0.28\textwidth]{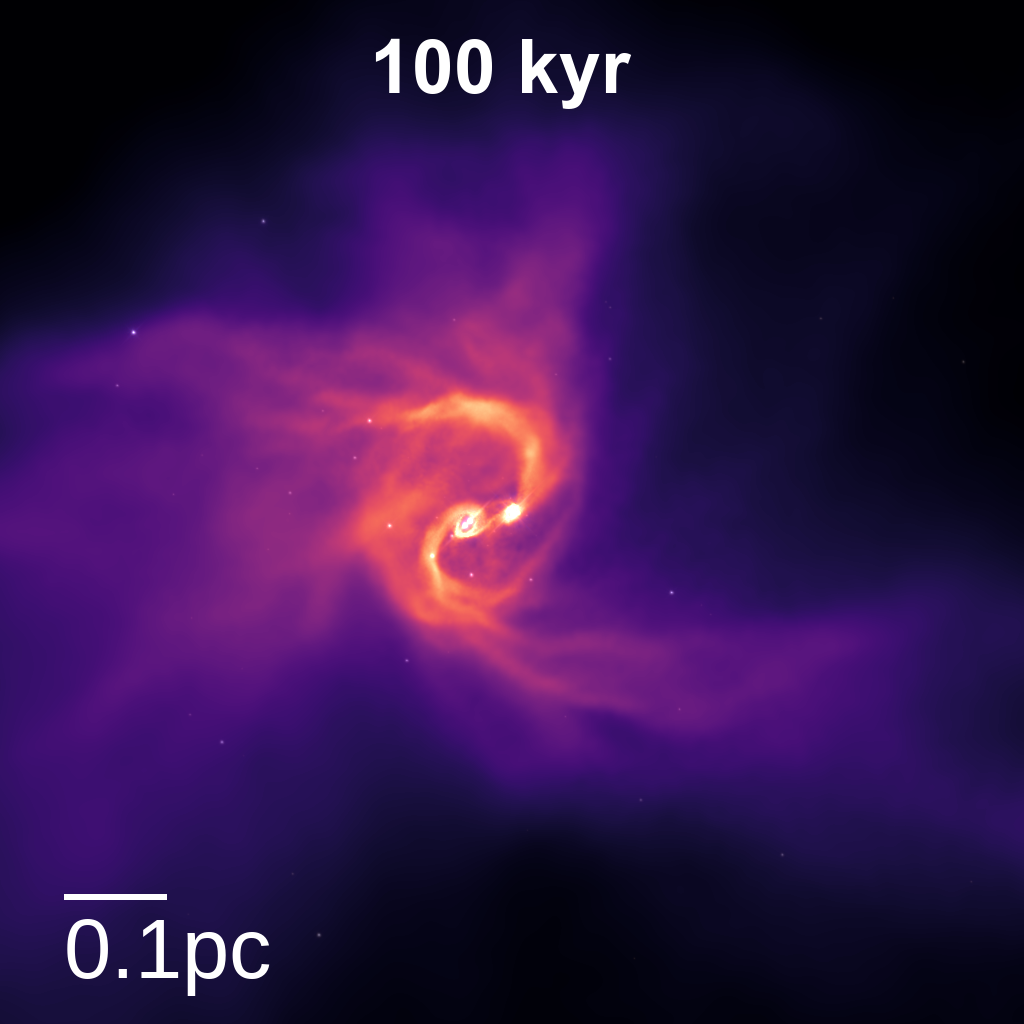}

\includegraphics[width=0.28\textwidth]{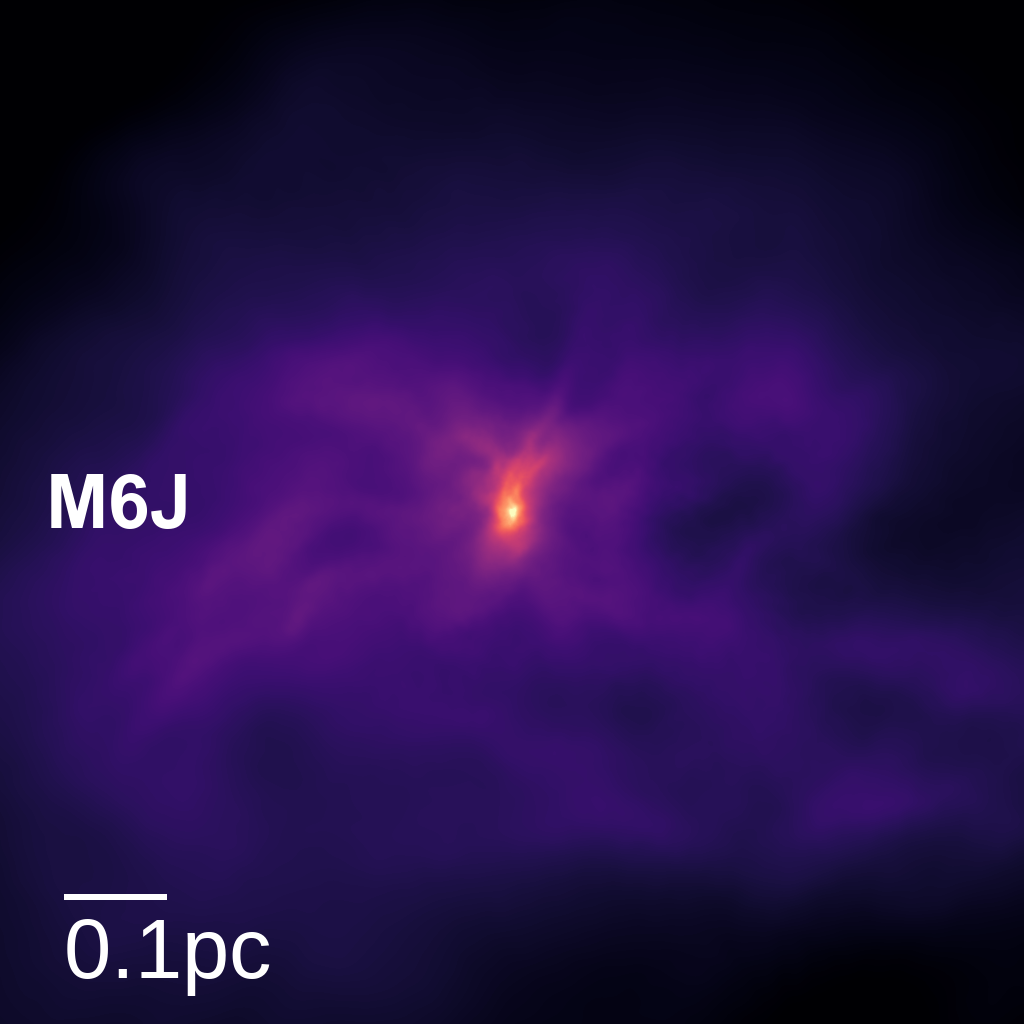}
\includegraphics[width=0.28\textwidth]{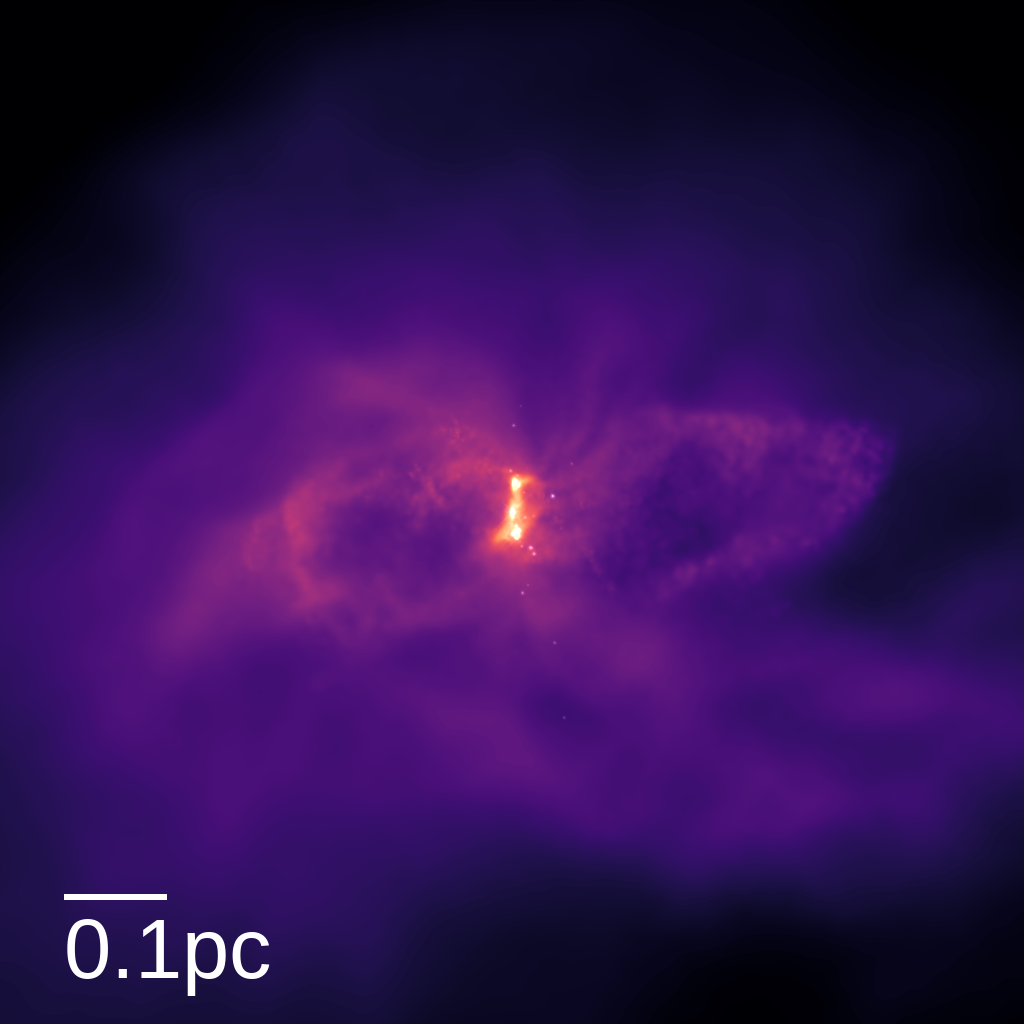}
\includegraphics[width=0.28\textwidth]{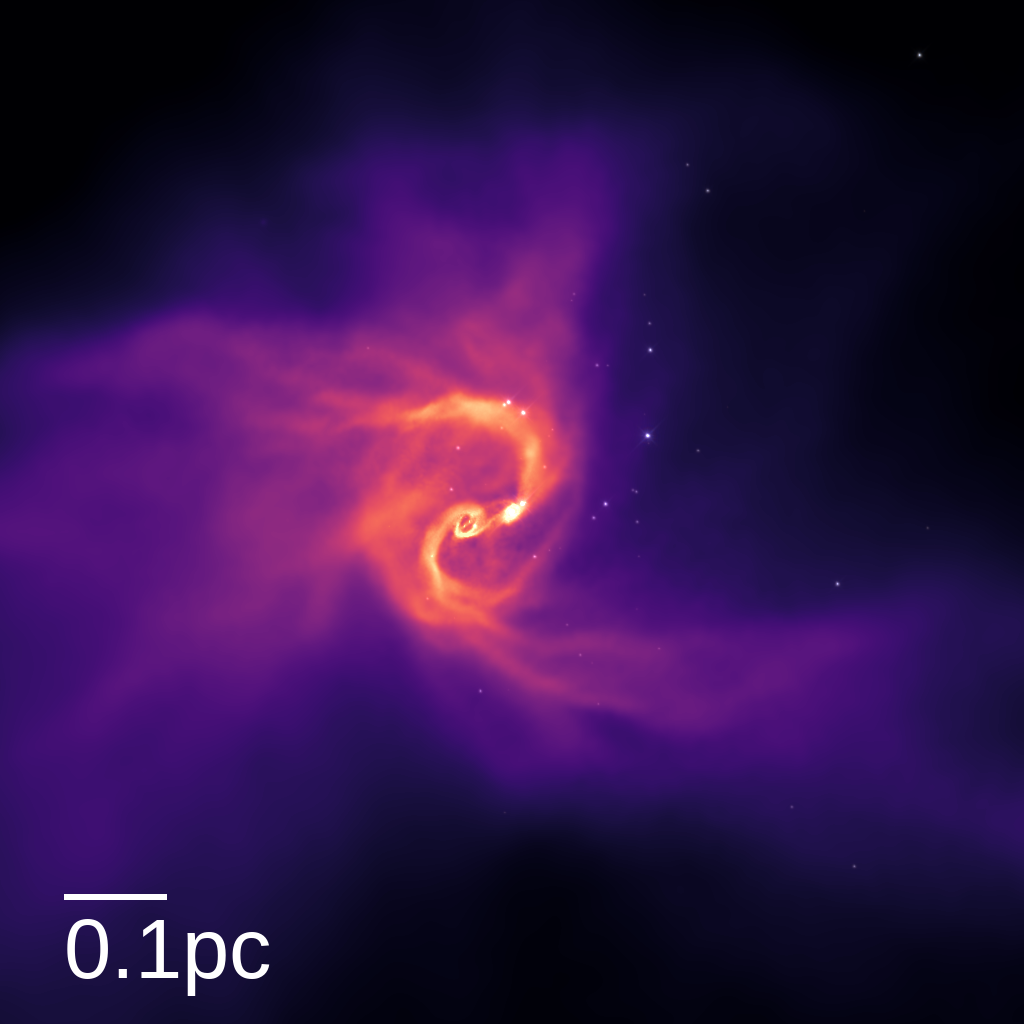}

\includegraphics[width=0.28\textwidth]{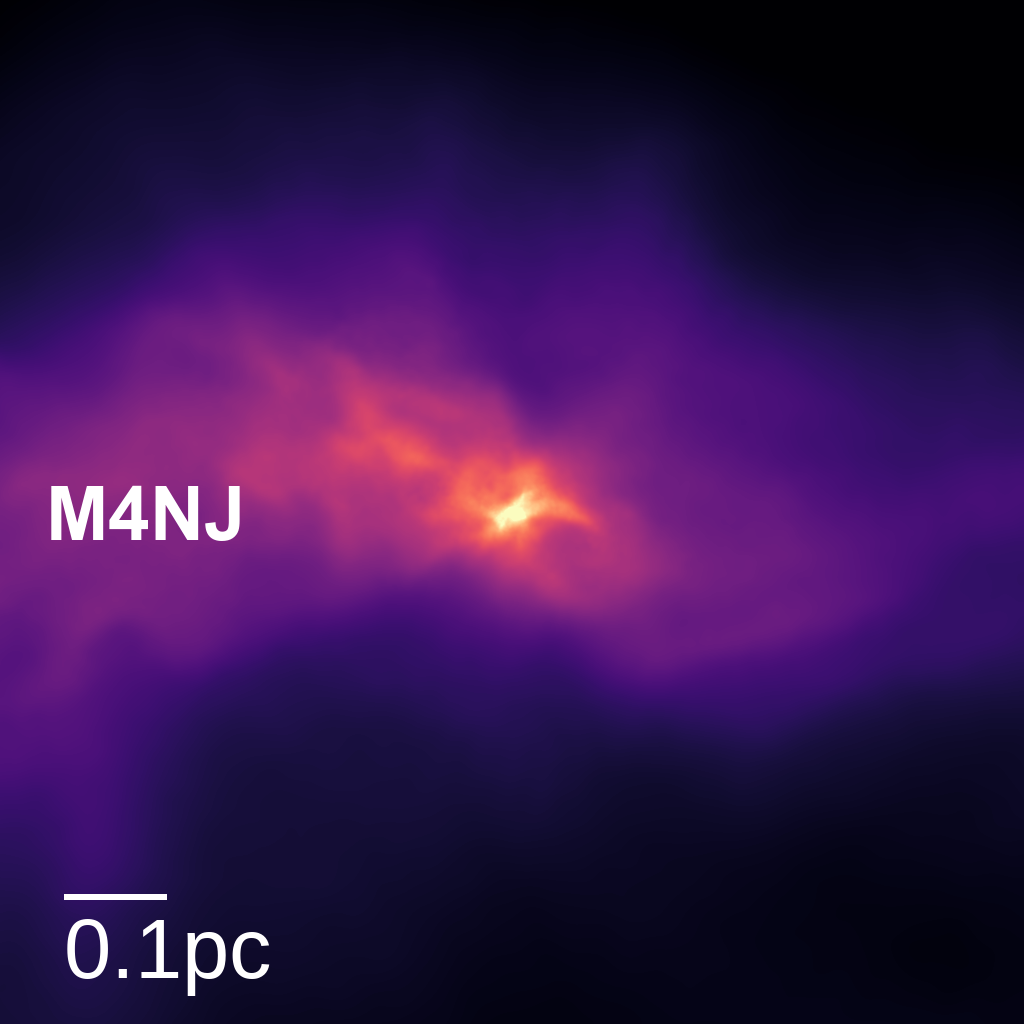}
\includegraphics[width=0.28\textwidth]{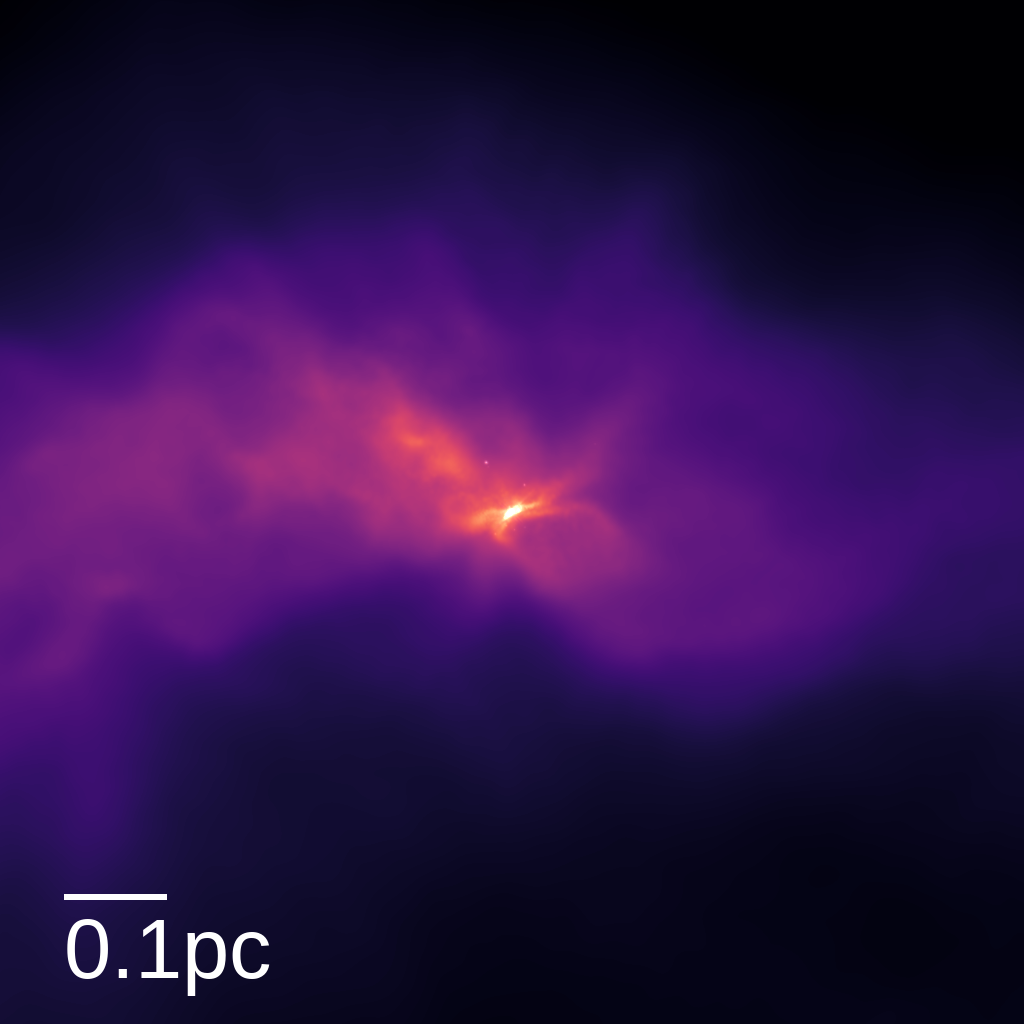}
\includegraphics[width=0.28\textwidth]{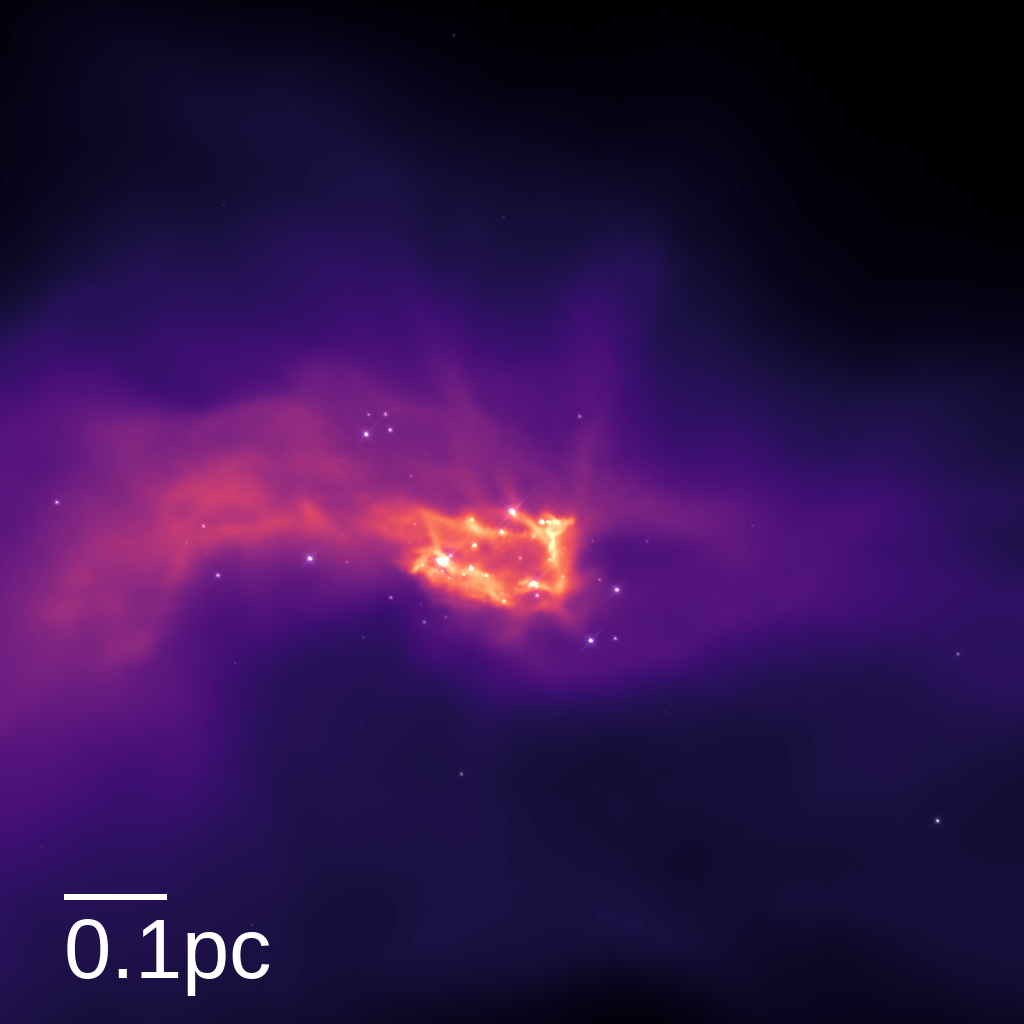}

\includegraphics[width=0.28\textwidth]{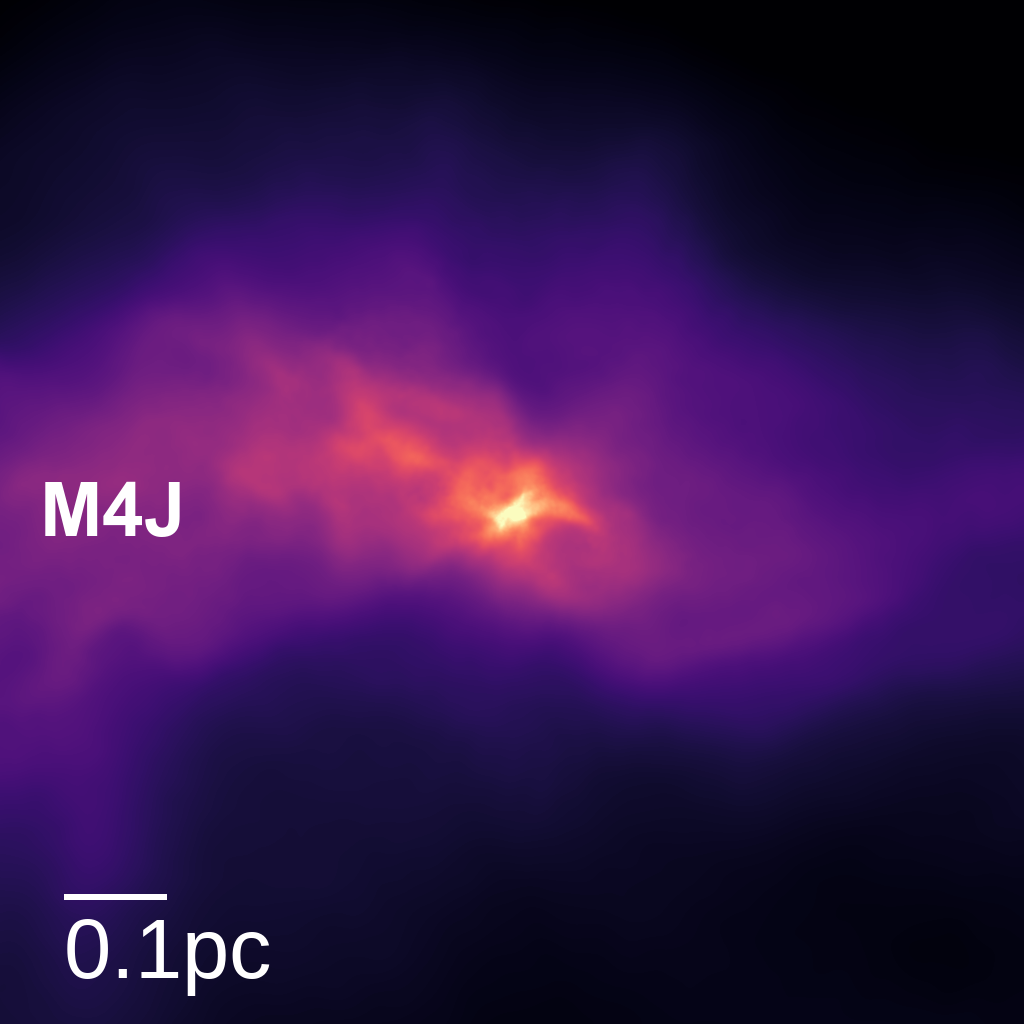}
\includegraphics[width=0.28\textwidth]{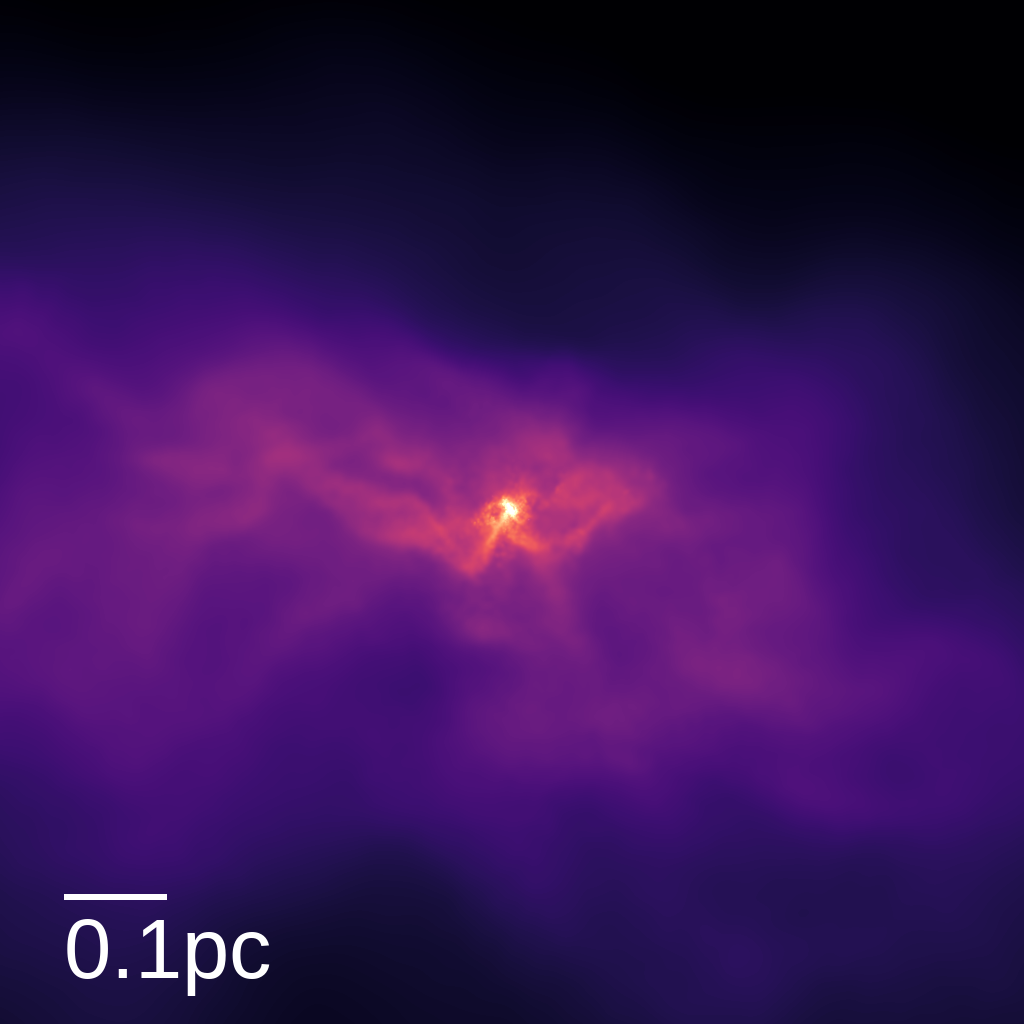}
\includegraphics[width=0.28\textwidth]{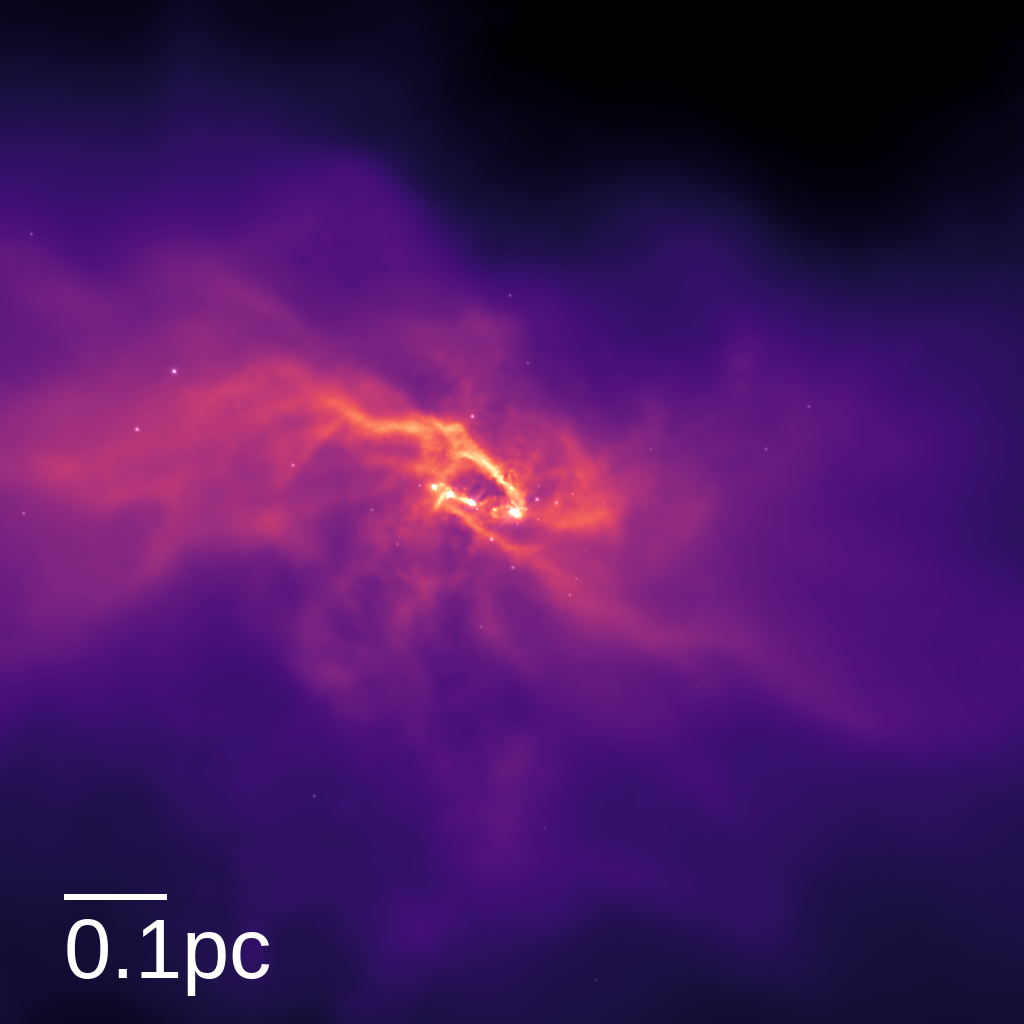}

\caption{Visualizations of the gas density in the star-forming region in our {\tt STARFORGE} simulations. The columns show (from left to right) the different systems $0$, $10$, and $100$ kyr after the formation of the first star. The rows (from top to bottom) are $10^{-6}$~Z$_\odot$ without (M6NJ) and with (M6J) jets, and $10^{-4}$~Z$_\odot$ without (M4NJ) and with (M4J) jets, as labelled in Table~\ref{Table:Runs}.}\label{fig:starforge}
\end{figure*}

In Figure~\ref{fig:starforge}, we show the central star-forming region of the evolved gas cloud (SIGO) in the {\tt STARFORGE} runs at three different times: at the formation of the first star $t_{\rm fs}$, $10$~kyr after $t_{\rm fs}$, and $100$~kyr after $t_{\rm fs}$ (columns from left to right). These visualizations are centered on the most massive star in each simulation. The low-metallicity $10^{-6}$~Z$_\odot$ run is shown in the top two rows (without jets, top row, and with jets, second row), and the $10^{-4}$~Z$_\odot$ run is shown on the bottom two rows (without jets, third row, and with jets, bottom row). For reference, we list the half-light radius and stellar mass surface density (using this half-light radius\footnote{This uses simulated masses, but note that direct comparison to observations is subject to observational uncertainty, especially because these stars would be UV-bright compared to a Kroupa or Salpeter IMF and short-lived due to their low metallicities and high masses.}) in Table~\ref{Table:Runs}. Because the high-mass and, therefore, high-luminosity stars are centrally concentrated in this initial burst of star formation, the stellar surface densities are very high, orders of magnitude above those of local star formation, and comparable to some of the highest known star cluster densities found in high-redshift stellar clusters observed with JWST \citep{Adamo+24}. Compared to the prior {\tt AREPO} simulations that these SIGOs are based on, these star clusters are both smaller and denser, owing to the higher resolution of the simulation, the shorter star formation timescales, and the presence of feedback. However, there are several caveats to these comparisons. Because of its young age, the present cluster is likely highly out of equilibrium. In addition, due to its young age and low gas mass, the cluster has a much lower stellar mass than the recent JWST discoveries.
 The cluster will expand on the 2-body relaxation timescale of order $10^5$ yrs because of its low mass, rapidly becoming less dense.

\begin{figure}[]
\centering
\includegraphics[width=0.45\textwidth]{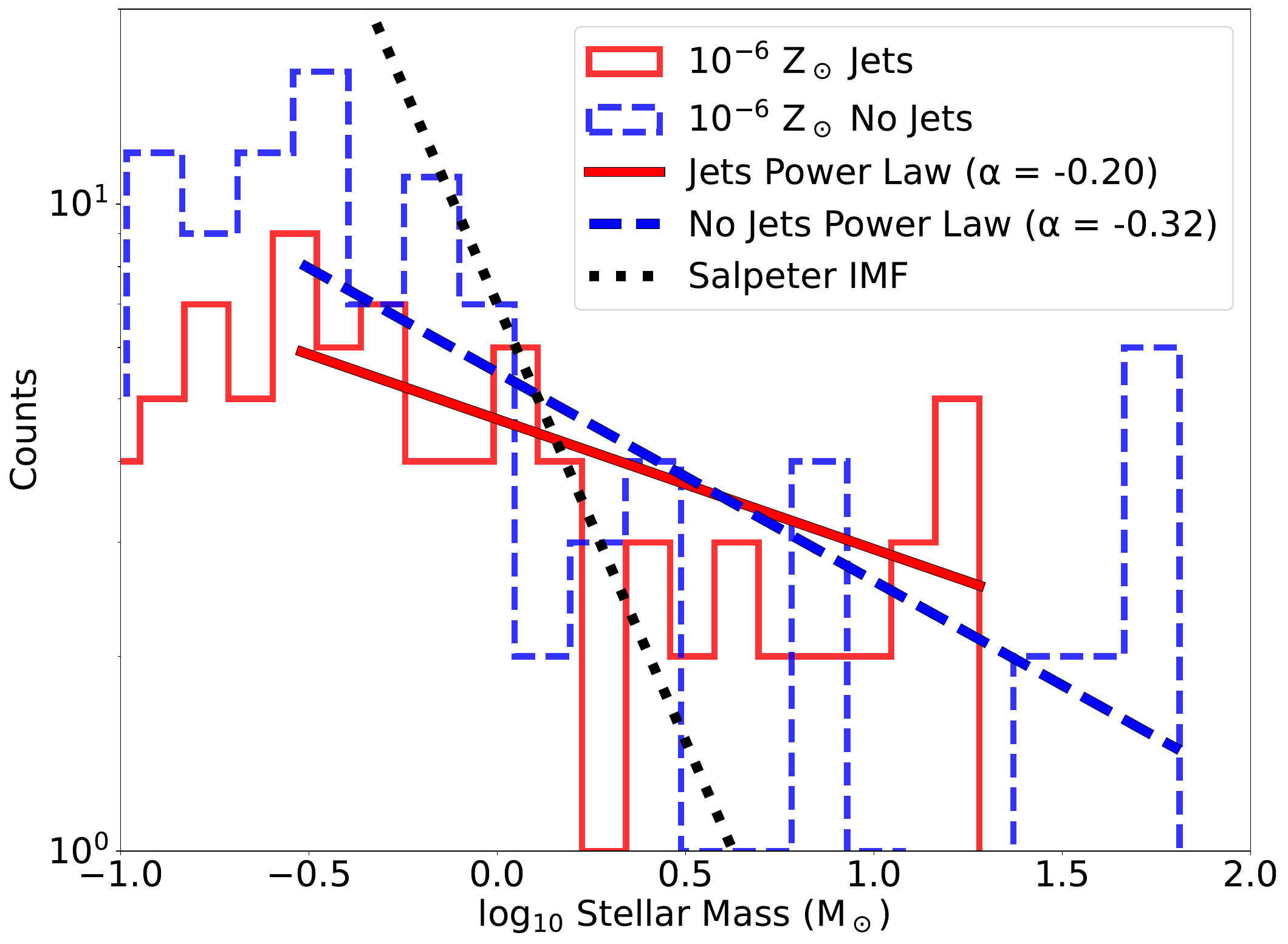}
\includegraphics[width=0.45\textwidth]{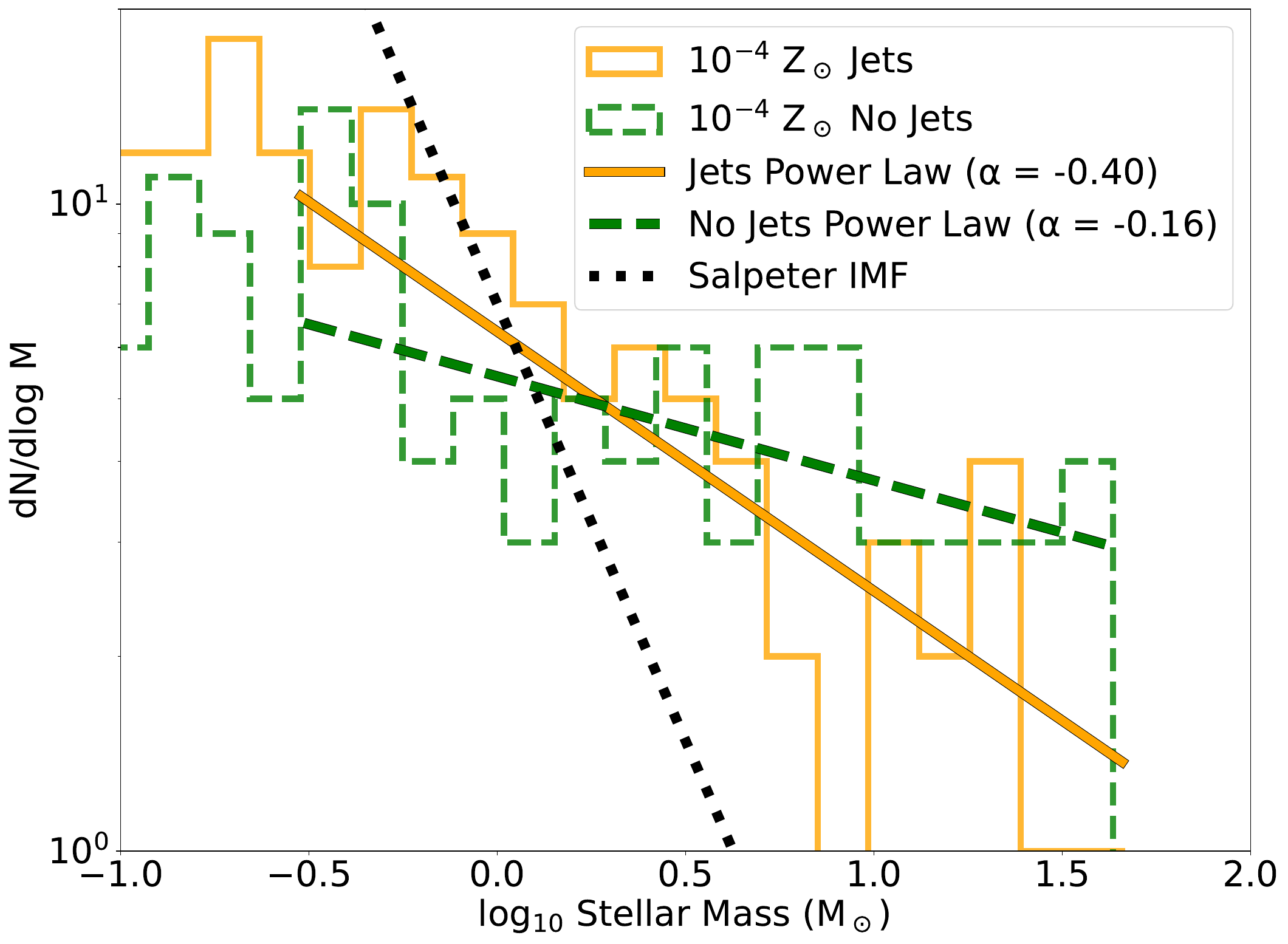}
\caption{The stellar mass function of SIGOs at different metallicities with and without feedback. The top panel shows simulations with $10^{-6}$Z$_\odot$, and the bottom panel shows the $10^{-4}$Z$_\odot$ cases. Dashed lines indicate simulations without jets, whereas solid lines indicate simulations with jets. The dotted lines represent a Salpeter mass function for comparison.}\label{fig:IMF}
\end{figure}

As is apparent from the rightmost column of Figure~\ref{fig:starforge}, the high-metallicity run has a tendency towards fragmentation on $0.01-0.1$~pc scales, regardless of feedback. This feature is more prominent in this case, owing to its enhanced cooling and lower temperatures on sub-pc scales. However, the level of fragmentation makes little difference to the total mass of stars formed in each run. 
In both runs without jets, the total stellar mass at $10^5$~yrs is about $\sim600$~M$_\odot$. 
On the other hand, the metallicity may have an impact on the maximum mass of the stars formed. Specifically, at $10^5$~yrs, the highest mass reached at the end of the simulation in run M6NJ is about $70$~M$_\odot$, compared to about $50$~M$_\odot$ in run M4NJ. The less fragmented accretion features in the lower-metallicity run permit substantially greater maximum masses of individual stars, even at these very low metallicities (as depicted in Figure~\ref{fig:IMF}). Assuming a star formation timescale of up to one Myr (after which, even if radiation has not quenched star formation, supernovae will), our results suggest a star formation efficiency in this SIGO of order $1\%$, with or without jets (though note that we do not account for the effects of radiative feedback, which could further limit the star formation efficiency, potentially to around the levels here at $10^5$~yrs). It is important to mention that as star-forming SIGOs reach $10^7$~M$_\odot$, this is a low-mass SIGO, which forms a low-mass and low-density star formation region. As the central density rises with the SIGO mass \citep{Lake+22}, it is likely that this is a lower bound on the star formation efficiency of high-mass SIGOs.

Given that our implemented jet feedback plays a minor role in the overall star formation efficiency in our simulated clouds, we can next ask whether it impacts the masses of the stars formed. In Figure~\ref{fig:IMF}, we show the stellar IMF of the $4$ simulations. The low-metallicity runs are shown in the top panel, and the $10^{-4}$~Z$_\odot$ runs are shown in the bottom panel. Runs with jets are indicated with solid lines, and runs without jets are indicated with dashed lines. We also report fits to the high-mass (M $\geq 0.3$~M$_\odot$, see Appendix~\ref{ssec:Convergence} for justification) end of the IMF of the form $dN/dM \propto M^{\alpha}$. For the low-metallicity runs, we find $\alpha=-0.32\pm0.12$ $(-0.20\pm0.05)$ for the case without (with) jets. For the $10^{-4}$~Z$_\odot$ runs, we find $\alpha=-0.16\pm0.05$ $(-0.40\pm0.05)$ for the case without (with) jets. Through a K-S test, we find no significant differences in $\alpha$ for the low-metallicity case ($p=0.35$ that the differences between the two distributions are due to random chance), but find support for significant differences in the high-metallicity case ($p=0.03$). In the high-metallicity case, we find a significantly steeper slope in the IMF with jets than without, owing to enhanced low-mass star abundances as the jets disrupt accretion flows.

In all cases, we find power-law stellar IMFs that are top-heavy, as is expected in Pop III star formation in other environments. For comparison, a Salpeter IMF is shown on each panel of Figure~\ref{fig:IMF}, indicating just how much more top-heavy these star clusters are than present-day stellar groups. This will likely result in low mass-to-light ratios in SIGOs: their enhanced luminosities could aid in detection (especially with the aid of gravitational lensing) in instruments such as JWST \citep{Lake+23}. If present, bipolar outflows do play a role: we find a stellar mass-to-light ratio in run M6NJ of $10^{-4} M_\odot/L_\odot$, $10\times$ lower than the $10^{-3} M_\odot/L_\odot$ in run M6J. This result is reflected to a lesser degree in runs M4NJ and M4J, with the mass-to-light ratio in M4J being 30\% higher than that in M4NJ. Also note that because the SIGO may form stars at later times, there is uncertainty in the final IMF--we can only provide an estimate at $100$~kyr. At later times, especially under the impact of escaping radiative feedback, high-mass star accretion is likely to be suppressed, steepening the IMF to a limited degree as accretion is quenched \citep{Chon+24}. 

Pop III star formation has been investigated before, both in zoom-in cosmological simulations \citep{Stacy+13,Hirano+14,Hirano+15,Prole+23} and in small cloud simulations \citep{Wollenberg+20,Jaura+22,Prole+22,2024ApJ...967L..28M}. The former often have either limited time evolution or mass resolution, while the latter is often envisioned as a small ($\lesssim 3000$~M$_\odot$) cloud within a bigger classical minihalo. 
For the first time, inspired by the DM-free structures seen in our cosmological simulations, we report a top-heavy IMF in a high-redshift, GC-like larger ($10^5$~M$_\odot$) cloud.  

\section{Conclusions}
\begin{itemize}
    \item Our simulations, which include models of thermochemistry and resolve fragmentation and individual stars, have demonstrated that early systems without dark matter are able to naturally form stars.
    \item We have shown that jet feedback has a minor impact on the low-metallicity high-mass IMF and does not substantially affect the star formation efficiency, although it may slightly steepen the final stellar mass function in SIGOs that are externally metal-enriched. 
    \item Additionally, we demonstrated that the process of star formation in SIGOs results in a top-heavy stellar mass function, producing bright stars with short lifetimes. This result is robust against potential external processes of metal enrichment in SIGOs, with even substantial levels of enrichment yielding top-heavy IMFs. These IMFs are much flatter than local low-metallicity young star clusters \citep[e.g.][]{kalari+18} owing to the near-pristine metallicities.
    \item SIGOs also form stars with very high stellar mass surface densities (even without invoking methods of reaching higher densities, such as ram pressure during accreting onto a minihalo), possibly comparable to anomalous observations at high redshift \citep{Adamo+24}. 
    
\end{itemize}

Given these results, formed clusters from SIGOs may be bright enough to be observed with JWST, if they form up to the Reionization era at $1\%+$ star formation efficiencies, especially with top-heavy IMFs \citep{Lake+23}. High-mass star-forming SIGOs may be able to form larger star clusters than the example SIGO here, resulting in UV-bright objects in JWST fields, which may be hosted within larger protogalaxies. These could manifest similarly to the star clusters already observed at high redshift, for example the Cosmic Gems clusters \citep{Adamo+24} as well as other observed high-z clusters and galaxies \citep{Messa+24,Vanzella+24, Fujimoto+25,Cullen+25}, which exhibit small sizes and low metallicities as would be expected of a DM-deficient star cluster through the SIGO formation mechanism. Clusters before Reionization, to be detected by future observations, may also originate from high-redshift SIGOs.

\acknowledgements

The authors thank the anonymous reviewer for their helpful comments and suggestions that have substantially strengthened the letter. W.L., S.N., C.W., B.B., F.M., and M.V. thank the support of NASA ATP grant No. 80NSSC20K0500 (19-ATP19-0020) and NASA  80NSSC24K0773 (APT 23-ATP23-0149), and the XSEDE and ACCESS AST180056 allocation, as well as the Simons Foundation Center for Computational Astrophysics and the UCLA cluster \textit{Hoffman2} for computational resources. S.N. thanks Howard and Astrid Preston for their generous support. B.B. also thanks the the Alfred P. Sloan Foundation and the Packard Foundation for support. M.V. acknowledges support through NASA ATP grants 16-ATP16-0167, 19-ATP19-0019, 19-ATP19-0020, 19-ATP19-0167, and NSF grants AST-1814053, AST-1814259,  AST-1909831 and AST-2007355. NY acknowledges financial support from JSPS International Leading Research 23K20035. C. E. W. acknowledges the support of the National Science Foundation Graduate Research Fellowship. This material is based upon work supported by the National Science Foundation Graduate Research Fellowship Program under grant No. DGE-2034835. Any opinions, findings, and conclusions or recommendations expressed in this material are those of the author (s) and do not necessarily reflect the views of the National Science Foundation. The research activities described in this paper have been co-funded by the European Union – NextGeneration EU within PRIN 2022 project n.20229YBSAN - Globular clusters in cosmological simulations and in lensed fields: from their birth to the present epoch. Numerical calculations were run on the TACC compute cluster “Frontera,” allocations AST21010, AST20016, and AST21002 supported by the NSF and TACC, and NASA HEC SMD-16-7592. This research is part of the Frontera computing project at the Texas Advanced Computing Center. Frontera is made possible by National Science Foundation award OAC-1818253

\appendix

\section{Convergence Tests}\label{ssec:Convergence}

\begin{figure}[]
\centering
\includegraphics[width=0.45\textwidth]{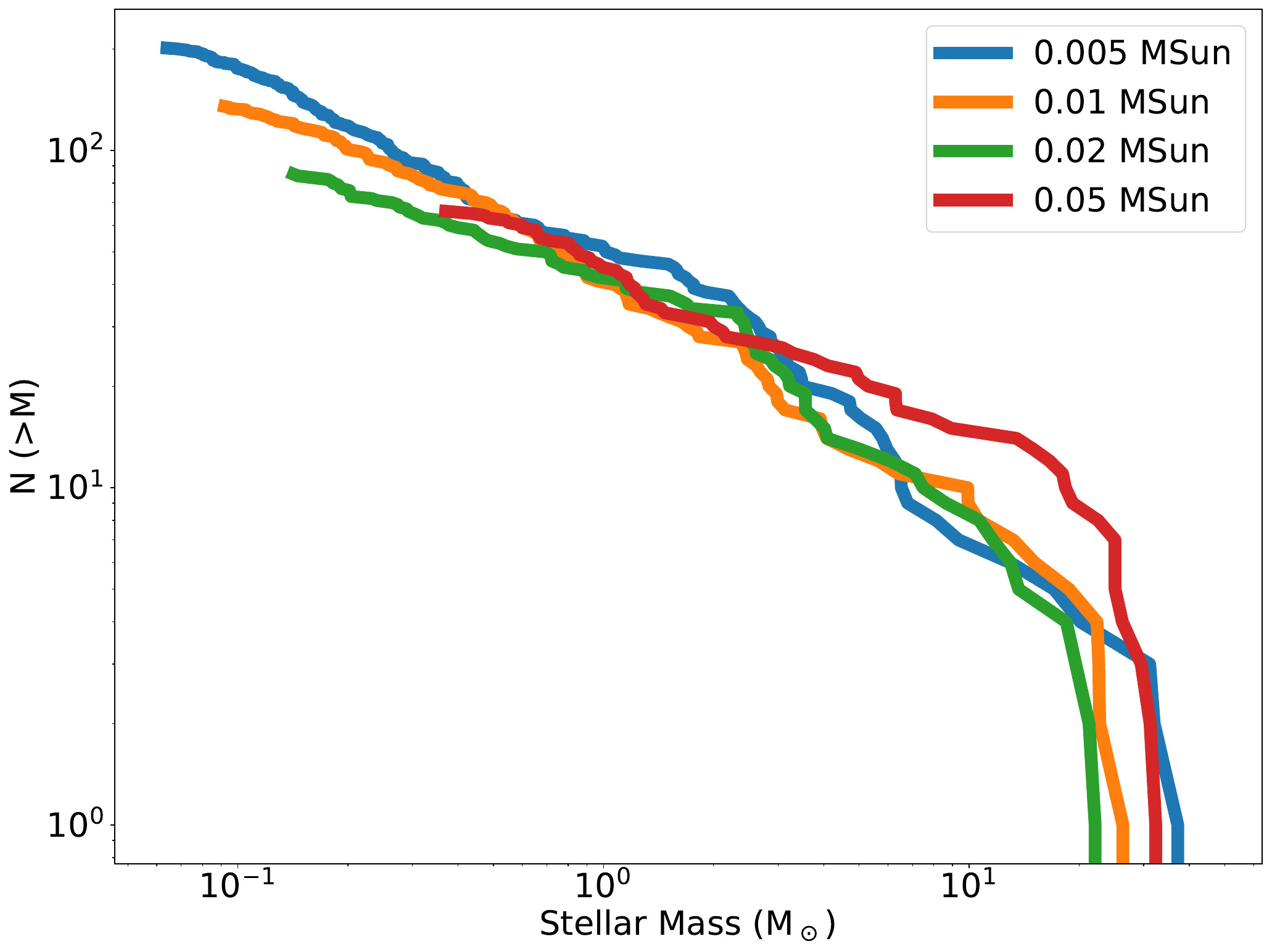}
\includegraphics[width=0.45\textwidth]{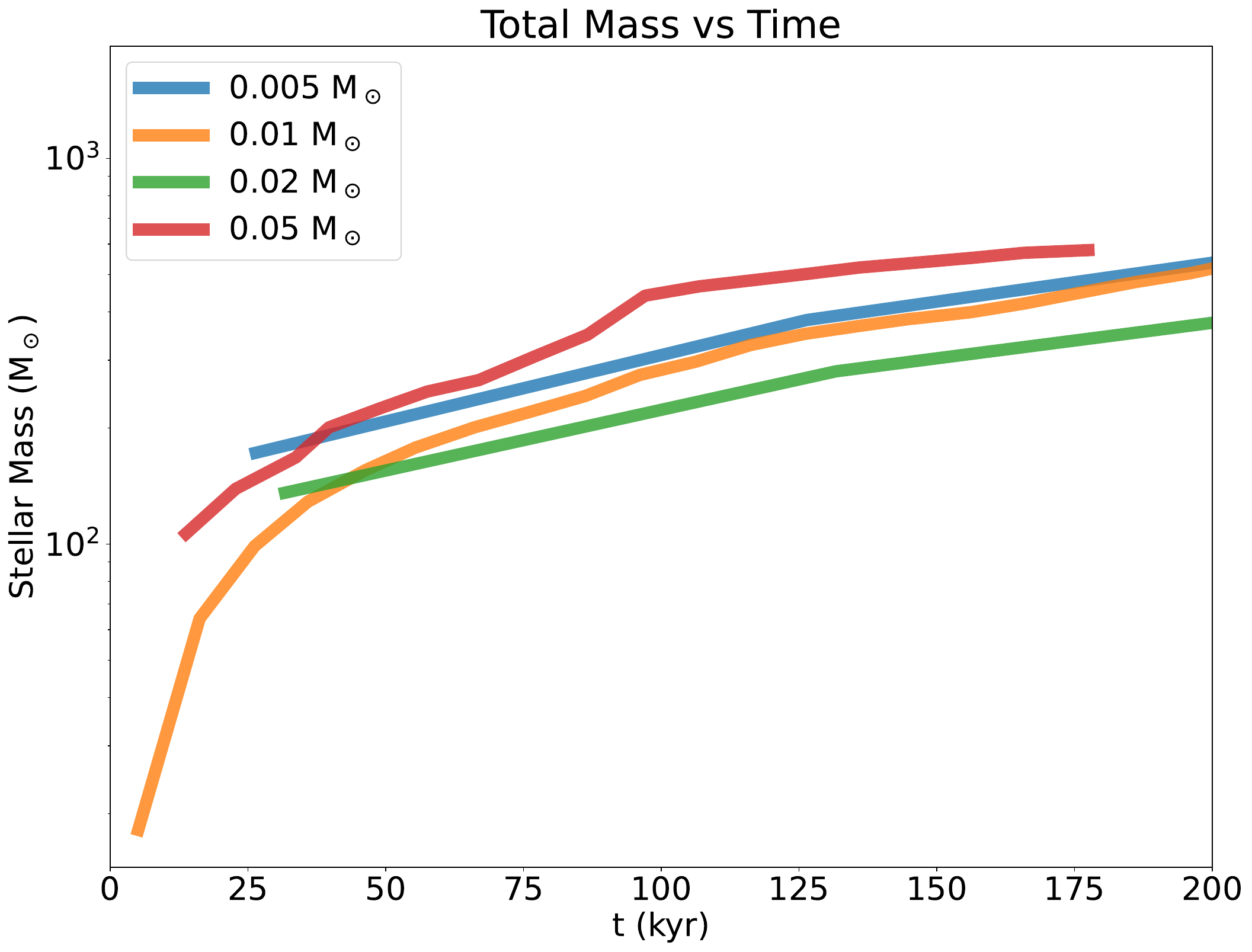}
\caption{Convergence tests of the IMF (left) and total stellar mass (right) in simulations of $0.005$~M$_\odot$ (blue), $0.01$~M$_\odot$ (orange), $0.02$~M$_\odot$ (green), and $0.05$~M$_\odot$ (red) resolution. The left panel depicts the IMF at $10^5$ years after the formation of the first star, and demonstrates convergence of the high-mass end of the IMF (stars resolved with at least $30$ gas particles in each run) for resolutions better than $0.02$~M$_\odot$.}\label{fig:convergence}
\end{figure}

As {\tt STARFORGE} has not previously been used to simulate Pop III or low-metallicity Pop II star formation, we run a suite of $4$ simulations at $0.005$, $0.01$, $0.02$, and $0.05$~M$_\odot$ resolution at Z$=10^{-4}$~Z$_\odot$ with jet feedback, allowing us to understand the effects of resolution on our results. These correspond to $6\times10^7$, $3\times10^7$, $1.5\times10^7$, and $6\times10^6$ particles, respectively. We tested the total mass and cumulative stellar IMF of each run, in order to check convergence of the results at decreasing resolutions. These tests are shown in Figure~\ref{fig:convergence}. The simulation with $0.005$~M$_\odot$ mass resolution is shown in blue, $0.01$~M$_\odot$ in orange, $0.02$~M$_\odot$ in green, and $0.05$~M$_\odot$ is shown in red. The left panel shows the cumulative stellar IMF $100$~kyr after the first star, and the right panel shows the total mass vs time following the formation of the first star.

In the left panel of this Figure, one can see that the stellar IMF is well-converged at mass resolutions below $0.02$~M$_\odot$, with the caveat that low-mass stars (those resolved with fewer than $30$ or so gas particles) are not converged, as their collapse is not well-resolved. Conclusions can, therefore, be drawn from the high-mass ($M>0.3$~M$_\odot$) IMF. Similarly, in the right panel, we see that the total mass in stars is well-converged at $100$~kyr between the $0.005$~M$_\odot$ and $0.01$~M$_\odot$ runs. Discrepancies occurring before this time may be attributed to our choice to track the time since the formation of the first star, as that is a somewhat stochastic event but made necessary by the relatively long pre-star-formation evolution of the cloud. These tests give us confidence that, in terms of the high-mass end of the IMF and total stellar masses, $0.01$~M$_\odot$ resolution is sufficiently accurate for the present suite of Pop III star formation simulations.

\section{Phase Diagram}\label{ssec:Phase}

\begin{figure*}[]
\centering
\includegraphics[width=0.67\textwidth]{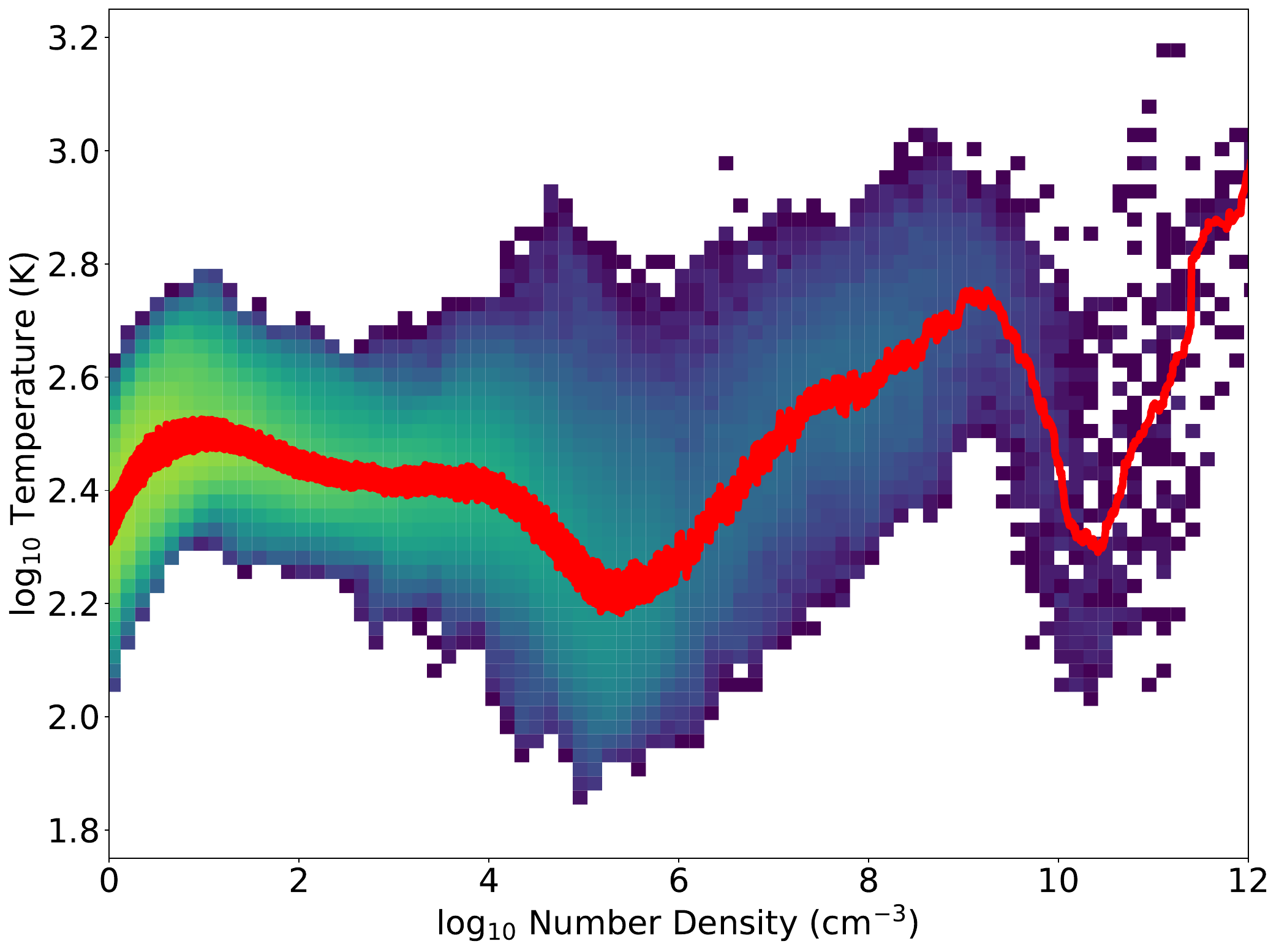}
\caption{The temperature and density of collapsing gas in our $10^{-6}$~Z$_\odot$ {\tt STARFORGE} simulation at the time of formation of the first star. The red line is a moving average of $100$ points nearest to the given number density and the yellow-violet background is a 2D histogram of all of the gas in our simulation at or above a number density of $1$~cm$^{-3}$.}\label{fig:phaseDiagram}
\end{figure*}

Owing to their lack of dark matter and thus relatively low masses, the early collapse process in SIGOs is characterized by relatively low virial temperatures of only a few hundred K. As seen in Figure~\ref{fig:phaseDiagram}, this results in a temperature peak in the early collapse of about $300$~K at n$\approx10$~cm$^{-3}$. At this point, molecular hydrogen cooling allows the gas to collapse while cooling to just under $200$~K by n$\approx10^5$~cm$^{-3}$, which is slightly lower than traditional Pop III star formation \citep{Klessen+23}, perhaps owing to the relatively long free-fall timescale in this system.

Following this initial collapse, characterized by lower temperatures than dark matter halos with a similar gas mass, the SIGO's collapse becomes more similar to that of molecular cooling minihalos. Protostellar collapse between about $10^5$~cm$^{-3} < $~n$_{\rm H} < 10^9$~cm$^{-3}$ is dominated in halos by PdV work from gas accreting onto the core, and not from heating through the halo potential, and in halos of a similar gas mass this causes heating to about $1000$~K by n$_{\rm H}\sim10^9$~cm$^{-3}$ \citep{Schauer+21,Klessen+23}. As expected, because this is not driven by dark matter, we see a similar heating process here. 

At n$_{\rm H}\sim10^9$~cm$^{-3}$, the 3-body reaction for molecular hydrogen formation becomes important, and as in halos, the gas becomes nearly fully molecular. This creates a brief period during which enhanced molecular hydrogen cooling permits the gas to cool sharply, followed by a return to PdV heating at n$_{\rm H}\sim10^{10}$~cm$^{-3}$ during the runaway collapse to protostellar formation. In this study, we form stars at about few$\times10^{12}$~cm$^{-3}$, and do not further trace the gas evolution.

\bibliography{myBib}{}
\bibliographystyle{aasjournal}



\end{document}